\documentstyle[aps,epsf,prl, preprint]{revtex}
\begin{document}
\title{Excitation spectrum of three dressed Bose-Einstein condensates}
\author{Zhong-Wen Ouyang and Le-Man Kuang*}
\address{ Department of Physics, Hunan Normal University, Changsha 410081, China} 
\maketitle

\begin{abstract}
  We study quantum dynamics of  three dressed Bose-Einstein condensates in a  high-Q
 cavity. The quasiparticle excitation spectrum of this system  is found  numerically.
  The stability of the  quasiparticle excitation is analyzed. It is shown that  
  there exist instabilities in the excitation spectrum.
\end{abstract}
 
]

  Recently,   multicomponent Bose-Einstein condensates (MBEC's) \cite{ho,law}  have 
  attracted much attention due to the recent experimental realizations of two and  
  three component Bose-Einstein condensates (BEC's) \cite{my,and}. MBEC's  offer 
  new degrees of freedom, which give rise to a rich set of new
 phenomena that do not exist in a single condensate. Especially, the interaction 
 of  BEC's  with light leads to fascinating effects including the extreme slowing 
 down of the speed of light \cite{hau} and  matter-wave four-wave mixing \cite{den}. 
 Goldstein, Wright and Meystre (GWM) \cite{gol} proposed an approach to produce 
 MBEC's, called dressed BEC's, inside a high-$Q$ multimode cavity,
 and investigated  the  quasiparticle excitation spectrum for the case 
 of two dressed BEC's which dress one-photon.
  The purpose of this letter is to study the  quasiparticle
 excitation spectrum for a system of three dressed BEC's which
 dress two photons. We will investigate quantum dynamics of the
 three dressed BEC's, and calculate numerically their elementary
excitations.

Consider  the GWM model of dressed BEC's [7], which consists of a
condensate which interacts with  two counter-propagating modes by
a high-Q ring cavity with   $n_1$ photons  being in the mode 1 and
$n_2$ photons being in the mode 2. We consider a state composing
$n=n_1+n_2$ photons and $N$ atoms which comprises the condensate
confined by an external trapping potential $U({\bf r})$.  The
state of the system can be written
$|\Psi_{N,n}\rangle=\sum_{n_1,n_2}\int dr_1 ...\int
dr_Nf_{n_1,n_2}({\bf r}_1,...,r_N,t)\Psi^{\dagger}({\bf
r}_1)...\Psi^{\dagger}({\bf r}_N)\\
\times|0,n_1,n_2\rangle$, where
$\Psi^{\dagger}({\bf r}_i)$ are the atomic field operators,
$f_{n_1,n_2}({\bf r}_1,...,r_N,t)$ is the many-particle
Schr\"{o}dinger wave function for the condensate, which can be
factorized as  a product of Hartree wave functions
$\phi_{n_1,n_2}$ representing   states which  the atoms occupy,
$f_{n_1,n_2}({\bf r}_1,...,r_N,t)=\prod_{i=1}^{N}\phi_{n_1,n_2}({\bf r}_i,t)$.

Let us consider such a trapping potential \cite{edw}  $U({\bf
r})=m\omega^2_{\perp}({\bf r}^2_{\perp}+\lambda^2z)^2$, where
$r=({\bf r}_{\perp}, z)$, ${\bf r}_\perp$ being the transverse
position coordinate, $\omega_{\perp}$ the transverse angular
frequency of the trap, and $\lambda=\omega_z/\omega_L$ is the
ratio of the longitudinal to transverse frequencies. For the cigar
structure condensate \cite{mew}, it is reasonable to assume that
$\lambda\ll 1$, and the transverse structure of the condensate is
determined by the ground state solution of the transverse
potential $\upsilon({\bf r}_{\perp})$. Then the Hartree wave
function can be expressed as 
 $\phi_{n_1,n_2}({\bf r},t)=\upsilon({\bf r}_{\perp})e^{-i\omega_{\perp}t}
 \psi_{n_1,n_2}(z,t)$.  
From the Hartree variational pronciple and  after integrating over the transverse
coordinate $r_{\perp}$,  one obtains Gross-Pitaeviskii equations (GPE's)  
\begin{eqnarray}
i\dot{\psi}_{n_1,n_2}(\xi,\tau)&=&H_L\psi_{n_1,n_2}(\xi,\tau)\nonumber \\ 
& & +\eta{\mid\psi_{n_1,n_2}(\xi,\tau)\mid}^2 \psi_{n_1,n_2}(\xi,\tau) \nonumber \\ 
& & + g[\sqrt{n_1(n_2+1)}\psi_{n_1-1,n_2+1}(\xi,\tau)  \nonumber \\
& & + \sqrt{(n_1+1)n_2}\psi_{n_1+1,n_2-1}(\xi,\tau)],
\end{eqnarray}
where ${\it H_L}=-\partial^2/\partial\xi^2+\xi^2/4$, with
$\xi=z/a_z$, $a_z=\sqrt{{\hbar}/{2{m} {\omega}_z}}$,
$\tau=\omega_zt$, 
$g$ and $\eta$ are the linear and nonlinear  coupling coefficients, respectively.

 The dressed BEC's are  the eigenstates of Eq.(1), and are quantum
 superposition states of states with different photon numbers
 $(n_1,n_2)$. We   consider the case of three dressed BEC's. In this case,
  only the states with (2,0), (1,1) and (0,2) are concerned.
The coupled GPE's (1) reduce to
\begin{eqnarray}
i \dot{\psi}_{2,0}(\xi,\tau) &=&H_L \psi_{2,0}(\xi,\tau)_+{\sqrt
2}g{\psi}_{1,1}({\xi,\tau})\nonumber
\\&+&\eta {\mid\psi_{2,0} (\xi,\tau)\mid}^2{\psi_{2,0}(\xi,\tau)},\nonumber
\\ i\dot{\psi}_{1,1}(\xi,\tau) &=&{\it
H}_L \psi_{1,1}(\xi,\tau)_+{\sqrt
2}g[{\psi}_{2,0}({\xi,\tau})\nonumber
\\&+&{\psi}_{0,2}({\xi,\tau})]+ \eta {\mid\psi_{1,1}
(\xi,\tau)\mid}^2 {\psi_{1,1}(\xi,\tau)}, \nonumber
\\ i \dot{\psi}_{0,2}(\xi,\tau) &=&{\it  H}_L
\psi_{0,2}(\xi,\tau)_+{\sqrt 2}g{\psi}_{1,1}({\xi,\tau})\nonumber
\\&+& \eta {\mid\psi_{0,2} (\xi,\tau)\mid}^2
{\psi_{0,2}(\xi,\tau)}.
\end{eqnarray}

 The exact analytic stationary solutions are generally not available for the 
 coupling GPE's of the dressed BEC's. We here consider their solutions under 
 the Thomas-Fermi approximation (TFA), in which the non-linear interaction 
 term dominates over the kinetic-energy term [9]. 
 Setting $\psi_{n_1,n_2}(\xi,\tau)={\it e^{-{\it i}\mu\tau}}
 {\theta_{n_1,n_2}(\xi)}$ with ${\mu}$ being the chemical potential, 
  under the TFA,  coupling  GPE's (2) reduce to
\begin{eqnarray}
{\mu\theta_{2,0}(\xi)}&=&{\frac{1}{4}}{\xi}^2\theta_{2,0}(\xi)+{\sqrt2}g{\theta_{1,1}({\xi})}
+\eta {\mid\theta_{2,0}(\xi)\mid}^2\theta_{2,0}(\xi), \nonumber \\
\mu\theta_{0,2}(\xi)&=&\frac{1}{4}{\xi}^2\theta_{0,2}(\xi)+\sqrt2g\theta_{1,1}({\xi})
+\eta \mid\theta_{0,2}(\xi)\mid^2\theta_{0,2}(\xi), \nonumber \\
{\mu\theta_{1,1}(\xi)}&=&{\frac{1}{4}}{\xi}^2\theta_{1,1}(\xi)+{\sqrt2}g[{\theta_{0,2}({\xi})}
+\theta_{2,0}(\xi)] \nonumber \\& &+\eta
\mid\theta_{1,1}(\xi)\mid^2\theta_{1,1}(\xi).
\end{eqnarray}

 For simplicity,  we consider only solutions of  two cases for
$\theta_{2,0}(\xi)$ and  $\theta_{0,2}(\xi)$: (a) Out-of-phase
solution with $\theta_{2,0}(\xi)=-\theta_{0,2}(\xi)$; (b) In-phase
solution with $\theta_{2,0}(\xi)=\theta_{0,2}(\xi)$.

(a)  {\it  Out-of-phase solution}.  In this case,
${\theta}_{2,0}({\xi}) =-{\theta}_{0,2}({\xi})$, from Eqs.(3) we
obtain the following dressed state nonzero solutions
\begin{equation}
{\theta}_{2,0}({\xi})=\sqrt{\frac{4\mu-\xi^2}{4\eta^2}},\hspace{0.5cm}
{\theta}_{1,1}({\xi})=0, \hspace{0.5cm} 4\mu\geq\xi^2.
\end{equation}

Let   ${\xi}_{\it m}$ be the boundary of the dressed condensates
 at which the  TFA solutions vanish. From Eq.(4) it follows that
 the chemical potential $\mu=\frac{1}{4}{\xi}_m^2$. The value of $\xi_m$
is determined by the normalization condition of the condensate
wave function
$\sum_{n_1,n_2}\int_{-\xi_m}^{\xi_m}d\xi\mid\theta_{n_1,n_2}(\xi)\mid^2=1$,
which leads to
 ${\xi}_m=[3{\eta}/2]^{1/3}$. So we get the following out-of-phase
 solution  

\begin{equation}
\mid\theta_{2,0}(\xi)\mid^2=\mid\theta_{0,2}(\xi)\mid^2=\frac{\xi_m^2-\xi^2}{4\eta},
\mid\theta_{1,1}(\xi)\mid^2=0.
\end{equation}

It is interesting to note that the out-of-phase solution of the
three dressed BEC's are the same as those of two dressed BEC's
found in Ref.\cite{gol} due to the vanishness of $\theta_{1,1}(\xi)$.
This reflects the fact that the three dressed BEC system can
transit to a two dressed BEC system.  Physically, this is
reasonable, because a three dressed BEC system naturally becomes a
two dressed BEC system when one of the three dressed condensates
vanishes.

(b) {\it In-phase solution.} In the case,
${\theta}_{2,0}({\xi})={\theta}_ {0,2}({\xi})$,  Eqs.(2) reduce to
two equations:
\begin{eqnarray}
(\mu-\frac{1}{4}\xi^2)\theta_{2,0}(\xi)&=&\sqrt2g\theta_{1,1}(\xi)
+\eta \mid\theta_{2,0}(\xi)\mid^2\theta_{2,0}(\xi), \nonumber \\
(\mu-\frac{1}{4}\xi^2)\theta_{1,1}(\xi)&=&2\sqrt2g\theta_{2,0}(\xi) \nonumber \\
& &+\eta \mid\theta_{1,1}(\xi)\mid^2\theta_{1,1}(\xi),
\end{eqnarray}
which  can be  solved  numerically. Assume $\xi_m$ to
be the boundary of the dressed condensates, from Eqs.(6) we then
obtain the chemical potential   ${\mu}^{\pm}={\pm}2{\it g}+
\xi_m^2/4$. Here the value of $\xi_m$ depends on the normalization
condition
$\sum_{n_1,n_2}\int_{-\xi_m}^{\xi_m}d\xi\mid\theta_{n_1,n_2}(\xi)\mid^2=1$.
For example, when   $g=7.29$ and $\eta=80$, we numerically  find that
$\xi^+_m=4.584$ and  $\xi^-_m=4.430$ coresponding to the chemical
potential being   ${\mu}^+$ and $\mu^-$, respectively.
   
We now   find the elementary excitations of the  
system by linearizing Eqs.(6) around the dressed state solutions:
$\psi_{n_1,n_2}(\xi,\tau)=e^{-i\mu\tau}[\theta_{n_1,n_2}(\xi)
+u_{n_1,n_2}(\xi)e^{-i\omega\tau}+ v_{n_1,n_2}^{\star}({\xi})
e^{i\omega\tau}]$, where$(n_1,n_2)=(2,0)$, $(1,1)$ and $(0,2)$, 
$u({\xi})$ and $v({\xi})$ represent small perturbations arround
the dressed state with energies ${\mu}\pm{\omega}$. Substituting
the expression of $\psi_{n_1,n_2}(\xi,\tau)$ into Eqs.(6), in first
order, we get the system of six equations for the linearized
perturbations $u({\xi})$ and $v({\xi})$:
\begin{eqnarray}
\omega
u_{2,0}(\xi)&=&[H_L+2\eta\mid\theta_{2,0}(\xi)\mid^2-\mu]u_{2,0}(\xi)
\nonumber \\ & & +\sqrt2gu_{1,1}(\xi)+\eta
\theta_{2,0}(\xi)^2v_{2,0}(\xi), \nonumber   \\
{\omega}u_{1,1}({\xi})&=&[H_L+2\eta{\mid\theta_{1,1}(\xi)\mid}^2
-\mu]u_{1,1}({\xi})  \nonumber \\
&&+\sqrt2g[u_{2,0}(\xi)+u_{0,2}(\xi)] +\eta
\theta_{1,1}(\xi)^2v_{1,1}(\xi), \nonumber  \\
 \omega u_{0,2}(\xi)&=&[H_L+2\eta\mid\theta_{0,2}(\xi)\mid^2
-\mu]u_{0,2}(\xi)  \nonumber
\\& & +\sqrt2gu_{1,1}(\xi) +\eta
\theta_{0,2}(\xi)^2v_{0,2}({\xi}), \nonumber \\
  -{\omega}v_{2,0}({\xi})&=&[H_L+2\eta\mid\theta_{2,0}(\xi)\mid^2 -\mu]v_{2,0}(\xi)  \nonumber
\\& &+\sqrt2gv_{1,1}({\xi}) +\eta \theta_{2,0}(\xi)^2u_{2,0}({\xi}), \nonumber  \\
-{\omega}v_{1,1}({\xi})&=&[H_L +2\eta{\mid\theta_{1,1}(\xi)\mid}^2
-\mu]v_{1,1}({\xi})  \nonumber
\\& &+\sqrt2g[v_{2,0}({\xi})+v_{0,2}({\xi})] +\eta
{\theta_{1,1}(\xi)}^2u_{1,1}({\xi}), \nonumber   \\
-{\omega}v_{0,2}({\xi})&=&[H_L +2\eta{\mid\theta_{0,2}(\xi)\mid}^2
-\mu]v_{0,2}({\xi})  \nonumber
\\& &+\sqrt2gv_{1,1}({\xi}) +\eta{\theta_{0,2}(\xi)}^2u_{0,2}({\xi}).
\end{eqnarray}

The normal modes of the above coupled equations are identical to
the elemantary excitations determined via Bogoliubov method. In
order to obtain the elementary excitations, we expand them in
terms of eigenfunctions of the trapping potential ${\it q}_{\nu}({\xi})$ as 
 \begin{eqnarray}
 u_{n_1,n_2}({\xi})&=&\sum_{\nu}{\it b}_{n_1,n_2}^{\nu}{\it q}_{\nu}({\xi}), \nonumber \\
 v_{n_1,n_2}({\xi})&=&\sum_{\nu}{\it c}_{n_1,n_2}^{\nu}{\it  q}_{\nu}({\xi}),
\end{eqnarray}
where  ${\it q}_{\nu}({\xi})$ satisfies $H_L {\it q}_{\nu}({\xi})=(\nu+\frac{1}{2}){\it q}_{\nu}({\xi})$.


It is  difficult to get an exact solution for the normal modes of
Eqs.(7). In  what follows, we give numerical solution  of them by
using the TFA. For the out-of-phase solution (4), we note that
 ${\mid\theta_{2,0}(\xi)\mid}^2=
{\mid\theta_{0,2}(\xi)\mid}^2\simeq{\mid\theta_{2,0}(\xi=0)\mid}^2
\approx \xi^2_m/4\eta$ for normal modes localized close to the
center of the trapping potential due to $\xi_m\gg1$.

Substitution of Eqs.(8) into Eqs.(7) gives a system of linear
equations for the coefficients $b_{n_1,n_2}^{\nu}$ and
$c_{n_1,n_2}^{\nu}$. The nonzero solution condition of the system
of linear equations gives rise to the excitation spectrum of the
elementary excitations. In the Fig. 1(a) we have plotted the
excitation spectrum in terms of the index $\nu$ of the linear
oscillator mode for the out-of-phase solution when $\eta=80$,
$g=7.29$ and $\eta_m=4.9324$. Fig. 1(b) is the  enlargement of the part 
of $-10<\omega^2<20$ for the  curve $c$. From Fig.1(a)  we see that in the
region of $0\leq \nu \leq 2$, there exists one type of  excitation
spectrum, which is stable due to $\omega^2>0$. In the region of
$\nu>2$ we can observe three types of  excitation spectrum. Two of
them indicated by curve $a$ and $b$ are stable  excitations. The third
one indicated by curve $c$ is instable in the region of $14<\nu <18$
since $\omega^2$ is negative. This is explicitly indicated  in Fig. 1(b) through 
the  enlargement of  the  curve $c$.


Similarly, we can numerically express the  excitation spectrum for
the in-phase solution   in terms of the index
$\nu$ of the linear oscillator mode.  In Fig. 2, we give rise to
the excitation spectrum for the in-phase solution when $\eta=80$,
$g=7.29$ and $\eta_m=4.932$, where we have set  that 
${\mid\theta_{2,0}(\xi)\mid}^2=
{\mid\theta_{0,2}(\xi)\mid}^2\simeq{\mid\theta_{2,0}(\xi=0)\mid}^2
\approx 0.0336$, ${\mid\theta_{1,1}(\xi)\mid}^2\simeq{\mid\theta_{1,1}
(\xi=0)\mid}^2\approx 0.0925$ for the branch ${\mu}^+$, and
$\mid\theta_{2,0} (\xi)\mid^2=\mid\theta_{0,2} (\xi)\mid^2\approx
0.04925$, ${\mid\theta_{1,1}(\xi)\mid}^2={\mid\theta_{1,1}(0)\mid}^2\approx
0.0775$ for the branch ${\mu}^-$, which lead to $\xi^+_m=4.584$ and $\xi^-_m=4.431$, 
 respectively. From Fig. 2(a), we see
that there exist different  excitation spectrum for three
different  regions of $\nu$. In the first region $0<\nu<8$, we can
find three kinds of  excitation spectrum, and  all of them  are stable.
In the second  region $8<\nu<13$,  there is only one  kind of
excitation spectrum   which  is  unstable (i.e., $\omega^2<0$). In the third region
$\nu>13$, there exist three kinds of  excitation spectrum. One of
them is stable, and other two are unstable. These unstable regimes are explicitly 
indicated  in  Fig. 2(b), which is the enlargement of the curves in the vicinity 
of the point A in Fig. 2(a).
From Fig. 2(c), we see that $\omega^2$ has two positive real number solutions and one pure imaginary 
solution which is not indicated in the figure. these means that there does not exist solutions 
of  $\omega^2<0$. Therefore, the excitation spectrum  of this case is always stable.

 {\it This work is supported in part by the National Climbing Program of China and NSF of  China,   the Excellence Young-Teacher 
Foundation of the Educational Department of China, and ECF and STF of Hunan Province.
The authors thank Dr.Changsong  Zhou for his help in preparing figures of the paper. }

* Corresponding author.

\begin{center}
{\bf Figure Captions}
\end{center}

FIG.1. (a) Excitation spectrum $\omega^2$ for the out-of-phase solution for 
           $\eta=80$, ${\it g}=7.29$, and  ${\eta}_m=4.932$. 
       (b) The enlargement of the part of $-10<\omega^2<20$ for the  curve $c$.

FIG.2. (a) Excitation spectrum $\omega^2$ for the in-phase solution with $\xi^+_m=4.584$. 
       (b) The enlargement of the curves in the vicinity of the point A in Fig.1(a).
       (c) Excitation spectrum $\omega^2$ for the in-phase solution with $\xi^-_m=4.431$.  Here we set $\eta=80$,  ${\it g}=7.29$. 
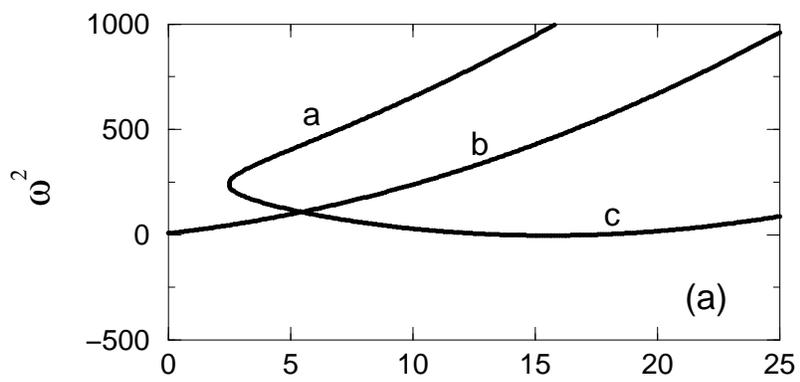
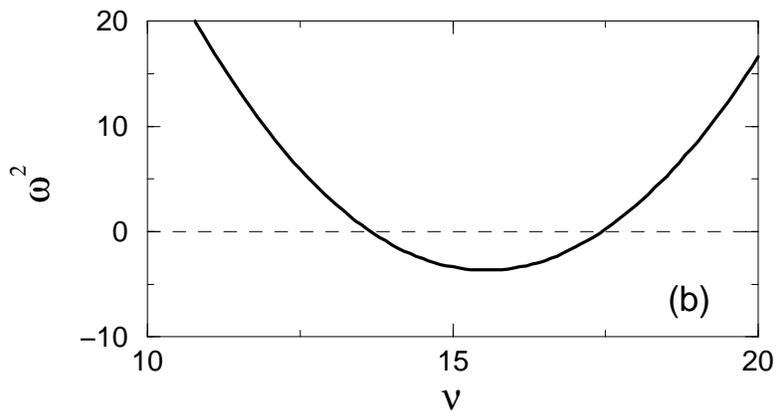
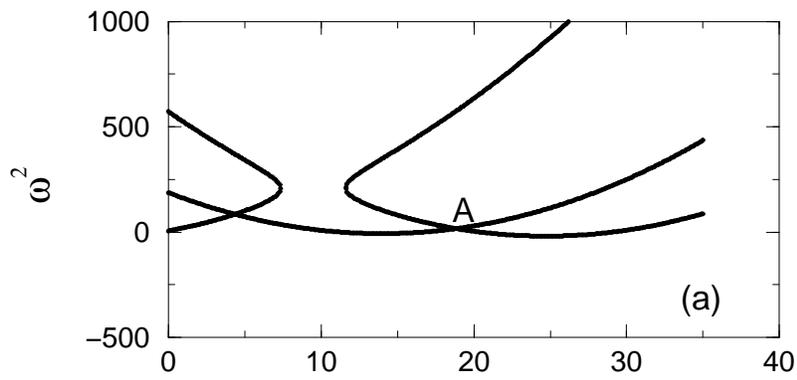
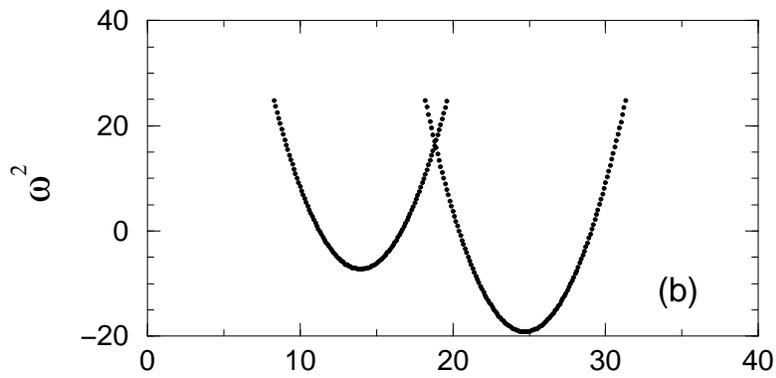
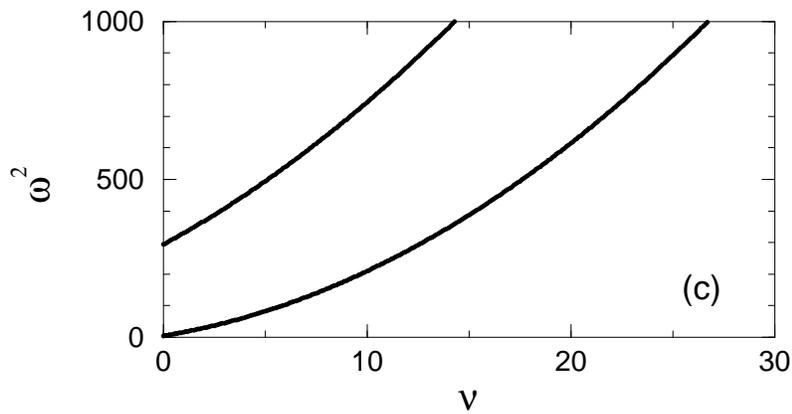
\end{document}